\documentclass[12pt]{article}
\pdfoutput=1

\usepackage{hyperref}

\usepackage{amsmath}
\usepackage{amssymb}
\usepackage{amsthm}
\usepackage[backend=bibtex8,hyperref=auto,style=numeric]{biblatex}
\usepackage[noabbrev]{cleveref}
\usepackage[letterpaper,left=1.2in,right=1.2in,top=1.2in,bottom=1.2in,marginparwidth=1.5in]{geometry}
\usepackage{graphicx}
\usepackage{mathabx}
\usepackage{mathtools}
\usepackage{microtype}
\usepackage{pgfplots}
  \pgfplotsset{compat=newest}
  \pgfplotsset{plot coordinates/math parser=false}
  \newlength\figureheight
  \newlength\figurewidth
\usepackage{siunitx}
\usepackage{tikz}
\usepackage{titling}
\theoremstyle{definition}
\newtheorem{definition}{Definition}
\AfterEndEnvironment{definition}{\noindent\ignorespaces}

\newtheorem{mathconjecture}{Math Conjecture}
\AfterEndEnvironment{mathconjecture}{\noindent\ignorespaces}
\newtheorem{physicsconjecture}{Physics Conjecture}
\AfterEndEnvironment{physicsconjecture}{\noindent\ignorespaces}

\AfterEndEnvironment{convention}{\noindent\ignorespaces}

\newtheoremstyle{defdotless}{}{}{}{}{\bfseries}{}{ }{}
\theoremstyle{defdotless}
\newtheorem{definitiondotless}[definition]{Definition}
\AfterEndEnvironment{definitiondotless}{\noindent\ignorespaces}
\newtheorem*{conventiondotless}{Convention}
\AfterEndEnvironment{conventiondotless}{\noindent\ignorespaces}

\DeclarePairedDelimiter\abs{\lvert}{\rvert} 
\DeclarePairedDelimiter\norm{\lVert}{\rVert} 

\newcommand\kTF{k_{\text{TF}}}
\newcommand\kTFtil{\tilde{k}_{\text{TF}}}
\newcommand\omegatrue{\omega_{\text{true}}}
\newcommand\RDM{R_{\text{DM}}}
\newcommand\tensor{\otimes}
\newcommand\vasy{v_{\text{asy}}}
\newcommand\vouter{v_{\text{outer}}}

\newcommand\Vtil{\tilde{V}}

\DeclareMathOperator\Ric{Ric}

\newcommand\R{\mathbb{R}}

\DeclareSIUnit\solarmass{\ensuremath{M_\Sun}}
\DeclareSIUnit\ly{ly}

\author{Hubert Bray, Andrew Goetz}
\title{Wave Dark Matter and the Tully-Fisher Relation}

\begin{document}

\maketitle
\begin{abstract}
We investigate a theory of dark matter called wave dark matter, also known as scalar field dark matter (SFDM) and boson star dark matter or Bose-Einstein condensate (BEC) dark matter, in spherical symmetry and its relation to the Tully-Fisher relation. We show that fixing the oscillation frequency of wave dark matter near the edge of dark galactic halos implies a Tully-Fisher-like relation for those halos. We then describe how this boundary condition, which is roughly equivalent to fixing the half-length of the exponentially decaying tail of each galactic halo mass profile, may yield testable predictions for this theory of dark matter.
\end{abstract}

\section{Introduction} \label{sec:intro}
Beginning in the 1970s, astronomers were surprised to discover that the stars in a typical spiral galaxy of baryonic mass $M_b$ are all orbiting the galactic center at roughly the same characteristic velocity $V$ 
\cite{rubin80,deblok08}, and furthermore that the quantity
\begin{equation} \label{eq:tullyfisher}
\kTF = \frac{M_b}{V^4} \approx 45\frac{M_\Sun}{(\text{km}/\text{s})^4}
\end{equation}
is a constant across galaxies \cite{tully77,mcgaugh10}. The latter relation is known as the (baryonic) Tully-Fisher relation. Since the rotational velocity of a galaxy depends on its mass, which is comprised mostly of dark matter, it is possible, if not probable, that the Tully-Fisher relation is intimately linked with the nature of dark matter.

Theories of dark matter abound. The most popular theory today is that dark matter is a Weakly Interacting Massive Particle, or WIMP, but if there is any connection between WIMPs and the Tully-Fisher relation, it has eluded discovery thus far. In this paper we investigate a theory we have termed \emph{wave dark matter} \cite{bray13b,bray12,bray13a,parry12b,parrythesis}. It has been investigated before under other names such as scalar field dark matter (SFDM) \cite{guzman00,magana12,suarez14} and boson star dark matter or Bose-Einstein condensate (BEC)  dark matter \cite{sin94,lai07,sharma08}. The difference in names comes from a difference in motivations, but the underlying equation is the Klein Gordon wave equation \eqref{eq:ekg2} for a scalar field.

Our main result is that fixing the oscillation frequency of wave dark matter near the edge of dark galactic halos implies a Tully-Fisher-like relation for those halos. Specifically, we require that
\begin{equation} \label{eq:sc1}
\omegatrue(\RDM+r_0)=\omega_0
\end{equation}
for some fixed $r_0,\omega_0$. Here $\omegatrue(r)$ is the frequency of the dark matter at radius $r$ and $\RDM$ is the radius of the dark matter halo, precisely defined later.  We comment that $r_0\ll\RDM$. In an upcoming paper \cite{goetz14} we show that this condition is one of a general class of ``Tully-Fisher boundary conditions'' that one can impose at the outer edge of dark halos, all of which produce Tully-Fisher-like relations. For example, another Tully-Fisher boundary condition, roughly equivalent to \cref{eq:sc1}, is to require that each wave dark matter halo mass profile has the same half-length for its exponentially decaying tail.

These results lead, given some assumptions, to testable predictions of the theory of wave dark matter. Specifically, if the conjectures in  \cref{sec:discussion} are correct, then we should be able to predict the total mass profile of a galaxy if we are given its baryonic mass profile.

\section{Modified Newtonian Dynamics} \label{sec:MOND}

We comment that there is another theory of dark matter, known as Modified Newtonian Dynamics or MOND \cite{milgrom83,famaey12}, which, while it has other issues, can claim to explain the flat rotation curves of spiral galaxies and the Tully-Fisher relation. Indeed, it was designed for this purpose. It is a bit of a misnomer to call it a theory of dark matter because, as its name suggests, instead of postulating the existence of extra matter which obeys the usual law of gravity, it modifies the law itself. In essence, whereas the combination of Newton's second law and law of gravity gives an acceleration due to gravity
\begin{equation*}
a = \frac{GM}{r^2},
\end{equation*}
MOND postulates
\begin{equation*}
a = \frac{\sqrt{GMa_0}}{r},
\end{equation*}
for an acceleration $a$ much less than a threshold acceleration $a_0$. The inclusion of the threshold acceleration is to leave solar system dynamics virtually unchanged. One immediately sees that for circular motion where $a=v^2/r$, we get $v=(GMa_0)^{1/4}$ (velocity independent of radius) and $M/v^4=(Ga_0)^{-1}$ (a Tully-Fisher relation), which seems promising. However, the theory has its own conflicts with data. One of the most problematic is that although MOND was created to get rid of the missing mass problem in galaxies, it has a missing mass problem at the level of clusters \cite{gerbal92,sanders99}. Even more problematic is the ``bullet cluster'', whose existence seems to demonstrate that dark matter exists in large quantities and can be separated from baryonic matter \cite{markevitch04}. Hence, MOND remains a minority viewpoint among astrophysicists. Even so, there is still the important question of why it works so well for spiral galaxies.

\section{Wave Dark Matter} \label{sec:WDM}
We now introduce the basics of wave dark matter so that we can describe our main result. Our main reference for the summary given here is \cite{bray13b}. Recall that in general relativity the fundamental object is a spacetime $N$ with a Lorentzian metric $g$ and stress-energy tensor $T$ satisfying the Einstein equation
\begin{equation} \label{eq:einstein}
G=8\pi T.
\end{equation}
(We use geometrized units throughout, with the gravitational constant and the speed of light set equal to $1$.) Here $G=\Ric-\frac{1}{2}Rg$ is the Einstein curvature tensor for $N$. It is well-known that the vacuum Einstein equation ($G=0$) can be obtained from a variational principle by requiring that the metric $g$ is a critical point of the Hilbert functional $\int_N R$ where $R$ is the scalar curvature. The connection $\nabla$ used to compute curvature is usually assumed to be the Levi-Civita connection. In \cite{bray13b} the author removed the assumption that the connection is the Levi-Civita one and investigated a general class of functionals which depend on the connection as well as the metric. The Hilbert functional is included in this class. It was found that deviation of the connection from the standard Levi-Civita connection could be described by a scalar function $f$ on the spacetime. Furthermore, requiring the metric $g$ and connection $\nabla$ to be critical points of the functionals led to the Einstein-Klein-Gordon (EKG) equations below.

\begin{subequations} \label{eq:ekg}
\begin{align}
G &= 8\pi\left( \frac{df\tensor d\bar{f}+d\bar{f}\tensor df}{\Upsilon^2} - \left(\frac{\abs{df}^2}{\Upsilon^2}+\abs{f}^2\right)g+T_b\right) \label{eq:ekg1} \\
\Box_g f &= \Upsilon^2 f \label{eq:ekg2}
\end{align}
\end{subequations}
The scalar function $f$ represents the dark matter, and we take it to be complex. We have added in a stress-energy tensor $T_b$ to allow for baryonic matter.  It turns out that we may set $T_b=0$ for the purposes of this paper, except for the stability conjectures stated in \cref{sec:discussion}. The constant $\Upsilon$ is a fundamental constant of nature yet to be precisely determined, though working values are given in \cite{magana12, bray13a}. For those who approach wave dark matter from a particle physics viewpoint instead of the geometric viewpoint described here, the fundamental constant is the mass $m$ of the dark matter particle. The relationship between $\Upsilon$ and $m$ is
\begin{equation}
m = \frac{\hbar\Upsilon}{c} = \SI{1.31e-22}{eV}\left(\frac{\Upsilon}{\SI{1}{\ly^{-1}}}\right).
\end{equation}

\section{Spherically Symmetric Static States} \label{sec:static}

We wish to investigate wave dark matter in the context of galaxies. The dark matter in galaxies is distributed in an approximately spherical halo, so we will take our spacetime metric to be spherically symmetric and write it in polar-areal form \cite{parry12a}:
\begin{equation}
g = -e^{2V(t,r)}\,dt^2 + \left(1-\frac{2M(t,r)}{r}\right)^{-1}\,dr^2 + r^2(d\theta^2 + \sin^2\phi\,d\phi^2).
\end{equation}
The functions $M$ and $V$ have natural Newtonian interpretations in the low-field limit as the total mass enclosed inside a ball of radius $r$ at time $t$ and the potential at radius $r$ and time~$t$. We remind the reader that we are using geometrized units so that the units for time, distance, and mass are all the same. In this paper we adopt the practice of measuring all quantities in years (which are the same as light-years). In \cref{table:masses,table:times,table:distances} at the end of this paper we give the values of some common astronomical units in geometrized units.

To make some formulas below more compact, we make the definition
\begin{equation}
\Phi(t,r) = 1-\frac{2M(t,r)}{r}.
\end{equation}
This term is approximately equal to $1$ in the low-field limit ($M\ll r$). Thus our spacetime metric is
\begin{equation}
g = -e^{2V}\,dt^2 + \Phi^{-1}\,dr^2 + r^2(d\theta^2 + \sin^2\phi\,d\phi^2). \label{eq:spacetime}
\end{equation}
We wish to solve the Einstein-Klein-Gordon (EKG) equations \cref{eq:ekg1,eq:ekg2}. A solution consists of a triple of functions $(M,V,f)$. The simplest solutions are the \emph{static states}, when $f$ is of the form
\begin{equation} \label{eq:staticstate}
f(t,r)=F(r)e^{i\omega t},
\end{equation}
with $F$ real and $M$ and $V$ functions of $r$ only. In this case the EKG equations simplify---see \cite{parry12b}---to the system of ODEs
\begin{subequations} \label{eq:ekgode}
\begin{gather}
M_r = 4\pi r^2\cdot\frac{1}{\Upsilon^2}\left[ \left(\Upsilon^2+\omega^2e^{-2V}\right)F^2 + \Phi F_r^2\right]\label{eq:ekgodeM} \\
\Phi V_r = \frac{M}{r^2} - 4\pi r\cdot\frac{1}{\Upsilon^2}\left[ \left(\Upsilon^2-\omega^2 e^{-2V}\right)F^2 - \Phi F_r^2 \right] \label{eq:ekgodeV} \\
F_{rr} + \frac{2}{r} F_r + V_rF_r + \frac{\Phi_r}{\Phi}F_r = \Phi^{-1}\left(\Upsilon^2-\omega^2e^{-2V}\right)F. \label{eq:ekgodeF}
\end{gather}
\end{subequations}
Note that the dependence on $t$ has disappeared, which is why solutions to these ODEs are called static states. Our notation for a solution will be
\begin{equation}
(\omega;M,V,F).
\end{equation}
Solutions can be found by numerical integration using a computer.

For initial conditions we take
\begin{align}
M(0)&=0 \label{eq:Minit}\\
V(0)&=V_0<0 \label{eq:Vinit}\\
F(0)&=F_0>0 \label{eq:Finit}\\
F_r(0)&=0. \label{eq:Frinit}
\end{align}
(These will be explained in a moment.) We define
\begin{definitiondotless}
\begin{align}
M_\infty = \lim_{r\to\infty} M(r) \\
V_\infty = \lim_{r\to\infty} V(r).
\end{align}
\end{definitiondotless}
It is important to note that we have the freedom to add an arbitrary constant $\Vtil$ to the potential function $V(r)$. Looking at the form of the ODEs \eqref{eq:ekgode}, we see that if $(\omega;M,V,F)$ is a solution, then so is $(\omega e^{\Vtil};M,V+\Vtil,F)$. This corresponds to just a rescaling of the $t$ coordinate by a factor of $e^{\Vtil}$ in the spacetime metric \eqref{eq:spacetime}. Thus, adding a constant to $V(r)$ amounts to a change of coordinates which does not affect the solution. In this paper, we use
\begin{conventiondotless}
\begin{equation} \label{eq:Vinf}
V_\infty = 0
\end{equation}
\end{conventiondotless}
for our solutions so that the metric \eqref{eq:spacetime} is asymptotic to Minkowski spacetime at infinity.

Now we can explain our initial conditions: \cref{eq:Minit} comes from the physical interpretation of $M(r)$ as the total mass inside the coordinate sphere of radius $r$. \Cref{eq:Vinit} is because of the convention \eqref{eq:Vinf}. \Cref{eq:Finit} is another convention: if $(M,V,F)$ is a solution, so is $(M,V,-F)$, and thus we might as well take $F(0)\geq 0$. We exclude $F(0)=0$ because this leads to the trivial solution. Finally, for regularity in spherical symmetry we must have $F_r(0)=0$; hence \cref{eq:Frinit}.

\begin{figure}[p]
\centering
\hrule
$n=0$\includegraphics[width=0.9\textwidth]{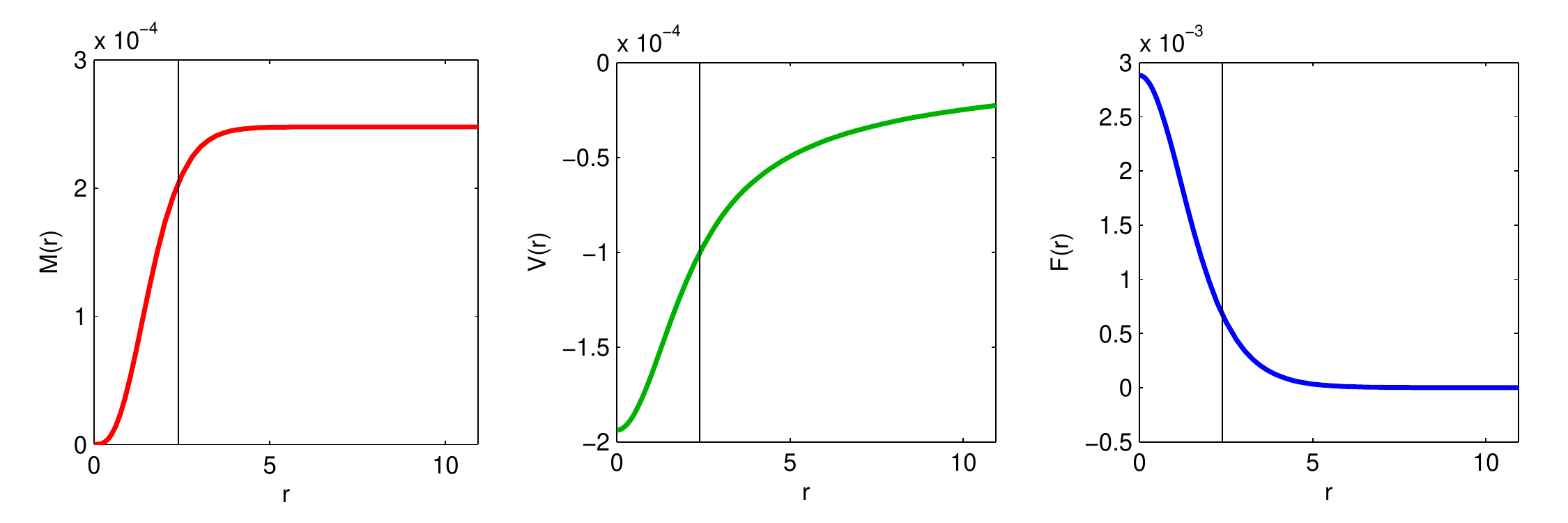}
\vspace{1pt}
\hrule
$n=1$\includegraphics[width=0.9\textwidth]{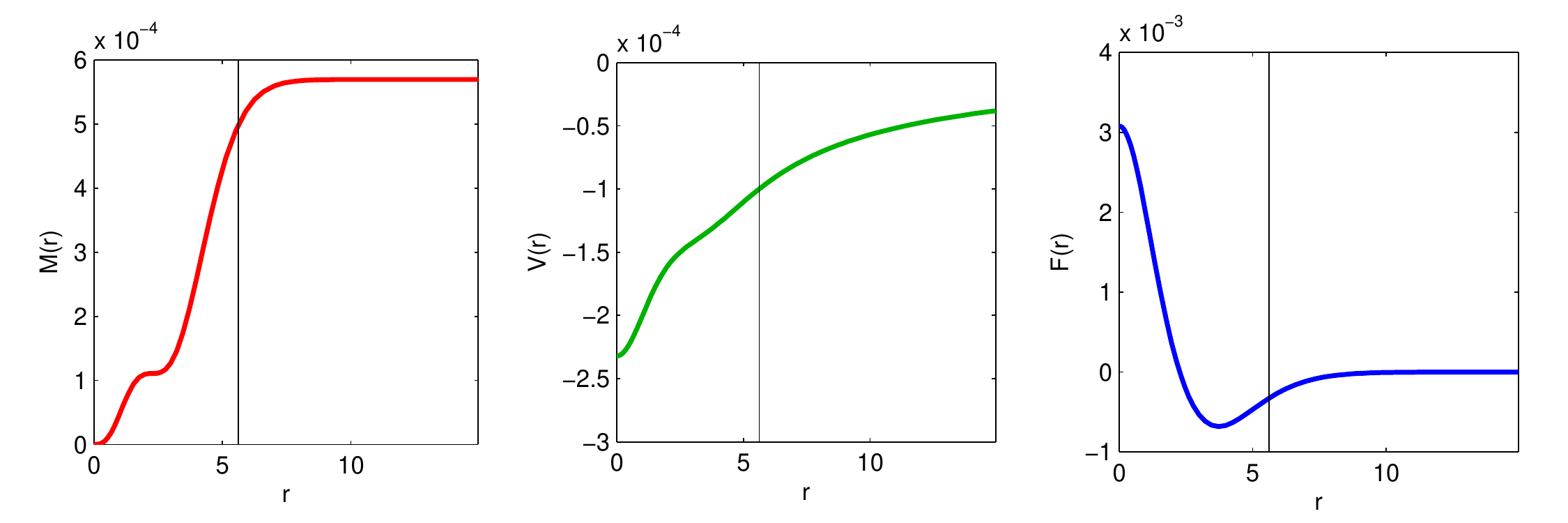}
\vspace{1pt}
\hrule
$n=2$\includegraphics[width=0.9\textwidth]{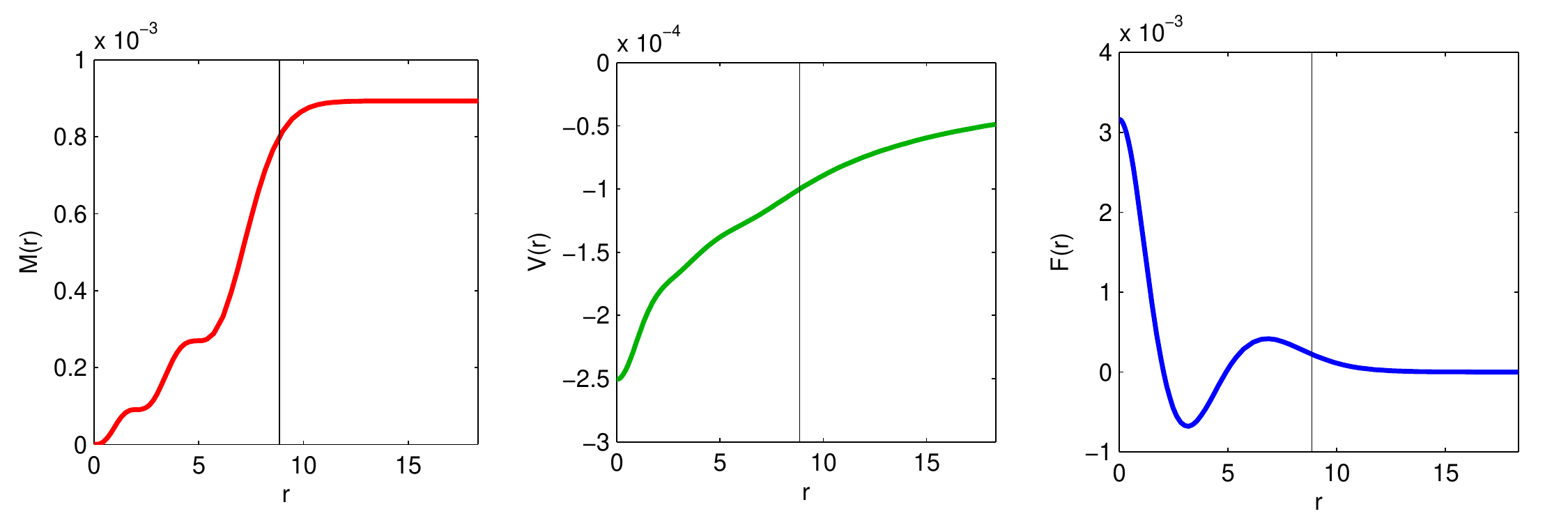}
\vspace{1pt}
\hrule
$n=3$\includegraphics[width=0.9\textwidth]{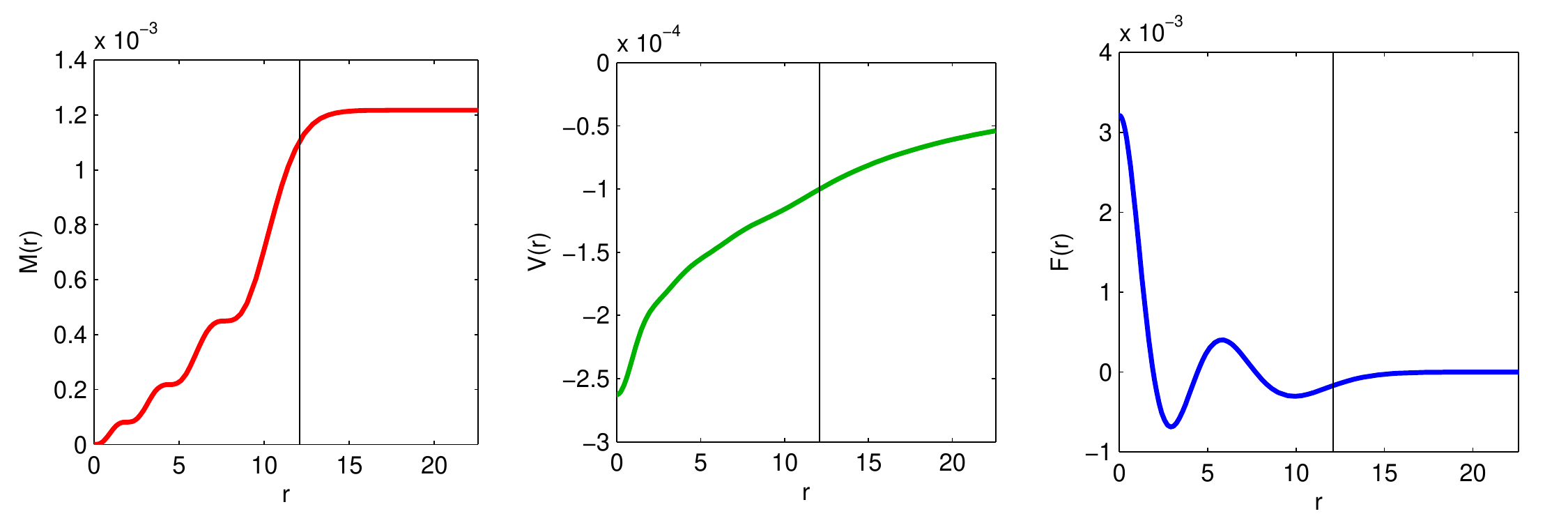}
\vspace{1pt}
\hrule
\caption{The ground state (top row) and first three excited states (second, third, and fourth rows) with $\Upsilon=100$ and $\omega=99.99$. The vertical black line in each plot marks the location of $\RDM$.}
\label{fig:n0123}
\end{figure}

Because we are describing finite mass systems, we require
\begin{equation}
M_\infty<\infty.
\end{equation}
Numerical experimentation shows that solutions having all the properties listed above come in the form of \emph{ground states} and \emph{excited states}. So that the reader can get a sense of the character of these solutions, in \cref{fig:n0123}
we have graphed the ground state and first three excited states with $\Upsilon=100$ and $\omega=99.9$. We see that for fixed $\Upsilon$ and $\omega$, there are countably many solutions corresponding to $n=0,1,2\ldots$ where $n$ is the number of zeros of $F$. To gain some insight into why this is the case, we consider the system \eqref{eq:ekgode} and make the following approximations:
\begin{gather}
e^{2V}\approx 1, \qquad \Phi\approx 1, \qquad \frac{V_r}{\Upsilon\norm{V}_\infty}\approx 0, \qquad \frac{\Phi_r}{\Upsilon\norm{V}_\infty}\approx 0 \label{eq:minkowskiapprox} \\
\frac{\omega}{\Upsilon} \approx 1, \qquad \frac{F_r}{\Upsilon\norm{F}_\infty}\approx 0. \label{eq:lowspeedapprox}
\end{gather}
The approximations in \eqref{eq:minkowskiapprox} may be interpreted as saying that the metric \eqref{eq:spacetime} is close to the Minkowski metric (the low-field limit), and the approximations in \eqref{eq:lowspeedapprox} may be interpreted as saying that the group velocities of wave dark matter are much less than the speed of light (the nonrelativistic limit). Applying all these approximations to the system \eqref{eq:ekgode} leads to the system
\begin{subequations} \label{eq:ekgodelimit}
\begin{align}
M_r &= 4\pi r^2\cdot 2F^2 \label{eq:ekgodeMlimit} \\
V_r &= \frac{M}{r^2} \label{eq:ekgodeVlimit} \\
F_{rr} + \frac{2}{r} F_r &= 2\Upsilon^2(1-\omega/\Upsilon+V)F. \label{eq:ekgodeFlimit}
\end{align}
\end{subequations}
As long as we are in the low-field, nonrelativistic limit, these systems are practically equivalent. See \cref{fig:lowfieldlimit}. Thus, to understand the system \eqref{eq:ekgode} in the low-field, nonrelativistic limit it suffices to understand the system \eqref{eq:ekgodelimit}.

From \cref{eq:ekgodeMlimit} we see that the quantity $2F^2$ is the mass-energy density at radius $r$. The second equation \eqref{eq:ekgodeVlimit} is familiar from Newtonian gravity. The third equation \eqref{eq:ekgodeFlimit} is the most interesting. The reader might recognize the left side as the expression for the Laplacian of a spherically symmetric function on $\R^3$. Let us consider for a moment the differential equation
\begin{equation} \label{eq:ekgodeosc}
F_{rr} + \frac{2}{r}F_r = k F.
\end{equation}
If we let $h(r)=rF(r)$, then the corresponding differential equation for $h$ is $h_{rr}=kh$. Thus the general solution to \cref{eq:ekgodeosc} is
\begin{equation} \label{eq:ekgFsol}
F(r) = \begin{cases} A\frac{e^{\sqrt{k}r}}{r} + B\frac{e^{-\sqrt{k}r}}{r} & k>0 \\ A + \frac{B}{r} & k=0 \\ A\frac{\sin(\sqrt{-k}r)}{r} + B\frac{\cos(\sqrt{-k}r)}{r}& k<0 \end{cases}
\end{equation}
Thus, on any interval where the expression $k(r)=2\Upsilon^2(1-\omega/\Upsilon+V(r))$ is roughly constant, the solution to \cref{eq:ekgodeFlimit} will look like one of the three solutions in \cref{eq:ekgFsol}. More explicitly, when $k(r)$ is negative, $F$ will exhibit oscillatory behavior, and when $k(r)$ is positive, $F$ will exhibit exponential behavior. Since $V$ is an increasing function, $k(r)$ increases with $r$. Remembering that one of our requirements on a solution $(M,V,F)$ is that $M_\infty<\infty$, looking at \cref{eq:ekgodeMlimit} we see that $F$ must exponentially decay after a certain point. Since we are also requiring $V_\infty=0$, we must take $\omega<\Upsilon$ so that $\lim_{r\to\infty} k(r) = 2\Upsilon^2(1-\omega/\Upsilon)>0$. On the other hand, $k(r)$ cannot be positive (i.e., $F$ cannot have exponential behavior) for all $r$; looking at \cref{eq:ekgFsol}, we see that this would be in contradiction with the requirements from spherical symmetry that $M_r(0)=F_r(0)=0$. Thus, the only situation consistent with our requirements is when $0<\omega<\Upsilon$, such that $k(r)$ begins negative and limits to $2\Upsilon^2(1-\omega/\Upsilon)$. The corresponding behavior of $F$ is to start out oscillating and then to switch over to exponential behavior. The switch occurs at the point where $k(r)=0$; we label this point $\RDM$ (see \cref{fig:n0123}) and view it physically as approximately the point where the dark matter halo ends, since $F$ decays exponentially thereafter.

\begin{figure}[tb]
\begin{minipage}{0.5\textwidth}
\centering
%
%
\begin{tikzpicture}

\begin{axis}[%
width=2.5in,
height=2.5in,
scale only axis,
xmin=0,
xmax=30.3671705142261,
xtick={18.5777082618857},
xticklabels={{$\RDM$}},
ymin=0,
ymax=0.002,
ytick={\empty},
ylabel={$M(r)$}
]
\addplot [color=red,solid,line width=2.0pt,forget plot]
  table[row sep=crcr]{%
0	0\\
0.0332356029814985	3.30393565761331e-09\\
0.0663421225062712	2.62159629016282e-08\\
0.10895010289847	1.15495094428698e-07\\
0.162720510647507	3.80784511390699e-07\\
0.227456179238181	1.02147402656692e-06\\
0.303635119169971	2.36100618381545e-06\\
0.392444218060836	4.87929229198223e-06\\
0.496509417141059	9.25769395723457e-06\\
0.61385252606422	1.59721894745456e-05\\
0.745613582304799	2.52892154203276e-05\\
0.899943700679824	3.74080152110055e-05\\
1.06715083195142	5.01033272625052e-05\\
1.23394732163963	6.04437057683893e-05\\
1.40679660101579	6.75241658663642e-05\\
1.58779317224012	7.09746464473203e-05\\
1.77698477022245	7.16977277663693e-05\\
1.93263515320482	7.18414577944492e-05\\
2.05855625240157	7.27987194258511e-05\\
2.19705178572113	7.56201243794328e-05\\
2.34489060807404	8.13453893264509e-05\\
2.49748483291217	9.0436781456343e-05\\
2.65124315064379	0.00010250950703349\\
2.80395156120509	0.000116489068122393\\
2.95481121804899	0.000131015029432632\\
3.10475265069429	0.000144849554418745\\
3.25369441013513	0.000156848014718459\\
3.40160877229764	0.000166237351433433\\
3.54833422671333	0.00017270766451209\\
3.69337957138205	0.000176432932759527\\
3.83566103092266	0.000178017136128541\\
3.97298214009595	0.000178355422067279\\
4.10048272783418	0.000178398752716061\\
4.2151311967687	0.000178839243171304\\
4.35837208841213	0.000180707454745647\\
4.52543204319736	0.000185701517968503\\
4.72199769153315	0.000196372701529952\\
4.97929366236527	0.000218059715639265\\
5.18207053143531	0.000239564777279029\\
5.38484740050535	0.00026233804426298\\
5.57244461508394	0.000282180720327505\\
5.7525659440315	0.000298341747158436\\
5.92764994265099	0.000310285963519315\\
6.09841743904801	0.000318033087302064\\
6.26465113237884	0.00032217713130623\\
6.42517397557998	0.000323754499652774\\
6.57730262068593	0.000324013994018763\\
6.71473645990409	0.00032409928348552\\
6.84589888915889	0.000324798378288252\\
7.00342056069896	0.000327335438981547\\
7.1829085752079	0.00033353632032721\\
7.38584553116686	0.000345737080811963\\
7.61980367396072	0.000366813902711448\\
7.92153565891401	0.000402721005451255\\
8.15964669564255	0.000434250113150965\\
8.39775773237109	0.000464581372801489\\
8.61845219343352	0.000488704533935465\\
8.82978686583184	0.000506462248678552\\
9.03453542231297	0.000518079897431494\\
9.23324206291973	0.000524437495892202\\
9.4250027683414	0.000526988559089664\\
9.60710571603504	0.000527483089842252\\
9.77301490517159	0.000527562849038937\\
9.92836471369826	0.000528374790315957\\
10.1115273063333	0.000531477711167144\\
10.3207857380744	0.000539322868169832\\
10.5565194885201	0.000555061969829905\\
10.8244021987855	0.000582430906586004\\
11.1455729293125	0.00062681108839906\\
11.4671092558459	0.000678876893624165\\
11.7886455823793	0.000731527503146344\\
12.0705025206859	0.00077268170178589\\
12.338241105806	0.000804081305770549\\
12.5989469574928	0.000825969791145792\\
12.854635630129	0.000839168433823304\\
13.1052722855428	0.000845432627698092\\
13.348739191266	0.000847247874936332\\
13.5791433493141	0.000847374515086316\\
13.7950699688249	0.000848166278891185\\
14.0336093147311	0.000852051525788519\\
14.3161821124093	0.000863611630140755\\
14.643636001985	0.000889757859568889\\
15.0279392334862	0.000940553664913736\\
15.5165374339546	0.00103538475504352\\
15.9676658662819	0.00114578079081227\\
16.4187942986091	0.00126709788542637\\
16.8367280257115	0.00137941838143683\\
17.2538906475726	0.00148380072632171\\
17.6849671573728	0.00157822287086016\\
18.146893201513	0.00166127222215547\\
18.6785447941452	0.00173351486559996\\
19.1851829139022	0.00178186046122615\\
19.6918210336593	0.00181434294761119\\
20.2112583885931	0.00183558089556507\\
20.755594772506	0.0018489306179238\\
21.2158585153598	0.00185560565227938\\
21.6761222582137	0.001859620133222\\
22.109941561054	0.0018618689839798\\
22.5252820119604	0.00186315496909781\\
22.9341137685184	0.00186390880140458\\
23.342398600116	0.00186435165916902\\
23.7541779439895	0.00186460979386899\\
24.1726540956997	0.00186475810062806\\
24.6006628908652	0.00186484164785438\\
25.0409308878338	0.00186488757667135\\
25.4962468002845	0.00186491209773101\\
25.9695999236272	0.00186492474660279\\
26.4643124905998	0.00186493101358671\\
26.9841833346068	0.00186493397452749\\
27.5336562722658	0.00186493529572617\\
28.1180218348194	0.00186493584411651\\
28.7436423216585	0.00186493604951054\\
29.4181105183013	0.00186493611332514\\
30.1499232862591	0.00186493612496156\\
30.3671705142261	0.001864936125092\\
};
\addplot [color=black,solid,forget plot]
  table[row sep=crcr]{%
18.5777082618857	0\\
18.5777082618857	0.002\\
};
\end{axis}
\end{tikzpicture}%
\end{minipage}%
\begin{minipage}{0.5\textwidth}
\centering
%
%
\begin{tikzpicture}

\begin{axis}[%
width=2.5in,
height=2.5in,
scale only axis,
scaled y ticks = false,
xmin=0,
xmax=30.3671705142261,
xtick={18.5777082618857},
xticklabels={{$\RDM$}},
ymin=-0.001,
ymax=0.0035,
ytick={\empty},
ylabel={$F(r)$}
]
\node at (axis cs:5,.001) [anchor=north west] {oscillating};
\node at (axis cs:20,.001) [anchor=north west] {decaying};
\addplot [color=blue,solid,line width=2.0pt,forget plot]
  table[row sep=crcr]{%
0	0.00327856487224789\\
0.0332356029814985	0.00327640824431574\\
0.0663421225062712	0.00326997910264732\\
0.10895010289847	0.00325545319208777\\
0.162720510647507	0.00322720156146554\\
0.227456179238181	0.003178842126406\\
0.303635119169971	0.00310265385826968\\
0.392444218060836	0.00298918456916862\\
0.496509417141059	0.00282575753991805\\
0.61385252606422	0.00260779857541694\\
0.745613582304799	0.00232964863364224\\
0.899943700679824	0.00197349213154801\\
1.06715083195142	0.0015699729164322\\
1.23394732163963	0.00116859788730362\\
1.40679660101579	0.00077167684364812\\
1.58779317224012	0.000392351548955964\\
1.77698477022245	4.83517400467923e-05\\
1.93263515320482	-0.000188164775833143\\
2.05855625240157	-0.000346666068655715\\
2.19705178572113	-0.000486324935310273\\
2.34489060807404	-0.000595627695978931\\
2.49748483291217	-0.00066690365728107\\
2.65124315064379	-0.0006988490546968\\
2.80395156120509	-0.000694804001126813\\
2.95481121804899	-0.000660337441606461\\
3.10475265069429	-0.000601033917502111\\
3.25369441013513	-0.00052254953609107\\
3.40160877229764	-0.000430405690492823\\
3.54833422671333	-0.000329896622374977\\
3.69337957138205	-0.000226096991086832\\
3.83566103092266	-0.000123900739513789\\
3.97298214009595	-2.81306427406701e-05\\
4.10048272783418	5.5910416356983e-05\\
4.2151311967687	0.00012592521826009\\
4.35837208841213	0.000204298424187268\\
4.52543204319736	0.000280894837513526\\
4.72199769153315	0.000348366423204584\\
4.97929366236527	0.000397805061246115\\
5.18207053143531	0.00040628428004793\\
5.38484740050535	0.000390116652237687\\
5.57244461508394	0.000356050379776085\\
5.7525659440315	0.000309127421290063\\
5.92764994265099	0.000253228785382282\\
6.09841743904801	0.000191872827812056\\
6.26465113237884	0.000128357555193946\\
6.42517397557998	6.58275466677929e-05\\
6.57730262068593	7.38902516045775e-06\\
6.71473645990409	-4.33581500137987e-05\\
6.84589888915889	-8.89810706836057e-05\\
7.00342056069896	-0.000139027052873765\\
7.1829085752079	-0.000188398356059791\\
7.38584553116686	-0.000232915361417326\\
7.61980367396072	-0.000267979322995201\\
7.92153565891401	-0.000286745639360501\\
8.15964669564255	-0.000281370177235227\\
8.39775773237109	-0.000260078630576165\\
8.61845219343352	-0.000228322000035125\\
8.82978686583184	-0.000189324135974881\\
9.03453542231297	-0.000145687965011081\\
9.23324206291973	-9.98271436410379e-05\\
9.4250027683414	-5.40469546929078e-05\\
9.60710571603504	-1.06349855942986e-05\\
9.77301490517159	2.78046279609392e-05\\
9.92836471369826	6.20732019312365e-05\\
10.1115273063333	9.95082541712363e-05\\
10.3207857380744	0.000137355486904156\\
10.5565194885201	0.000172585777976788\\
10.8244021987855	0.000202016659915702\\
11.1455729293125	0.000221672408790372\\
11.4671092558459	0.000224565381654044\\
11.7886455823793	0.000212116157420825\\
12.0705025206859	0.000190421728865263\\
12.338241105806	0.000162332586457953\\
12.5989469574928	0.000129727187139395\\
12.854635630129	9.43524497458497e-05\\
13.1052722855428	5.78894137146767e-05\\
13.348739191266	2.20361702356143e-05\\
13.5791433493141	-1.13021412233971e-05\\
13.7950699688249	-4.13081578973614e-05\\
14.0336093147311	-7.23963825933512e-05\\
14.3161821124093	-0.000105650258721645\\
14.643636001985	-0.000138466675150176\\
15.0279392334862	-0.000168337916890567\\
15.5165374339546	-0.000192492714086825\\
15.9676658662819	-0.000201856265052708\\
16.4187942986091	-0.000200543761929997\\
16.8367280257115	-0.0001917688082581\\
17.2538906475726	-0.000177758074850968\\
17.6849671573728	-0.000159762567152237\\
18.146893201513	-0.000138560416556115\\
18.6785447941452	-0.000113968813037258\\
19.1851829139022	-9.20251408925479e-05\\
19.6918210336593	-7.25482676913559e-05\\
20.2112583885931	-5.5600600841758e-05\\
20.755594772506	-4.11684399107916e-05\\
21.2158585153598	-3.14326031793821e-05\\
21.6761222582137	-2.36849342881517e-05\\
22.109941561054	-1.79381304568496e-05\\
22.5252820119604	-1.36181737803495e-05\\
22.9341137685184	-1.02958798407498e-05\\
23.342398600116	-7.72594176034108e-06\\
23.7541779439895	-5.73995521297925e-06\\
24.1726540956997	-4.21302349880298e-06\\
24.6006628908652	-3.04850490246409e-06\\
25.0409308878338	-2.1697955094716e-06\\
25.4962468002845	-1.51536398983384e-06\\
25.9695999236272	-1.03548313496879e-06\\
26.4643124905998	-6.89925688412711e-07\\
26.9841833346068	-4.46243317496398e-07\\
27.5336562722658	-2.78411956794146e-07\\
28.1180218348194	-1.65704856690997e-07\\
28.7436423216585	-9.16808061506311e-08\\
29.4181105183013	-4.31454735408111e-08\\
30.1499232862591	-8.80357845674182e-09\\
30.3671705142261	-1.64549371683279e-10\\
};
\addplot [color=black,solid,forget plot]
  table[row sep=crcr]{%
18.5777082618857	-0.001\\
18.5777082618857	0.0035\\
};
\end{axis}
\end{tikzpicture}%
\end{minipage}
\caption{A typical fifth excited state ($n=5$) demonstrating the location of $\RDM$ (see \cref{def:rdm}). We have omitted the plot of the potential $V(r)$. To the left of $\RDM$, $F(r)$ exhibits oscillatory behavior. To the right of $\RDM$, $F(r)$ exhibits exponentially decaying behavior. Note also that almost all the dark matter is contained in the region $r\leq\RDM$, or in other words, $M(\RDM)\approx M_\infty$.}
\label{fig:RDM}
\end{figure}
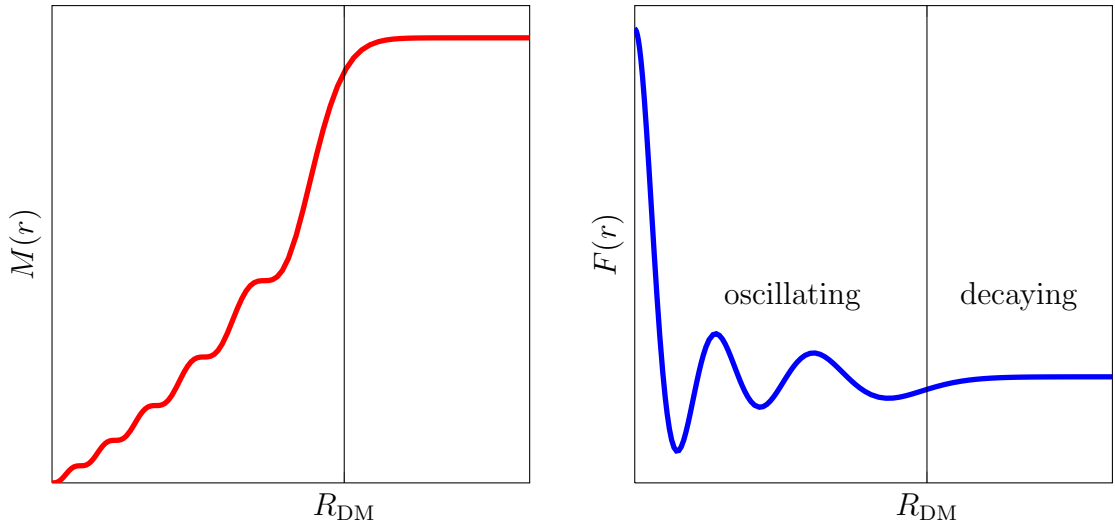

\begin{figure}[p]
\centering
\hrule
$\omega=99$\includegraphics[width=0.89\textwidth]{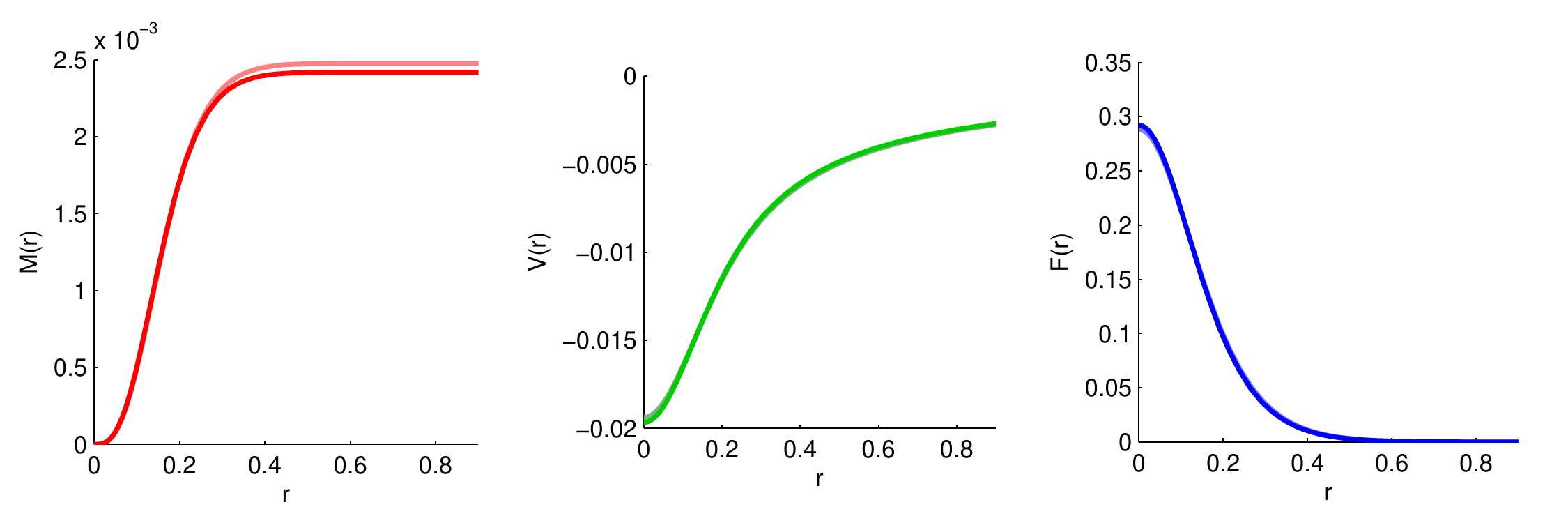}
\vspace{1pt}
\hrule
$\omega=95$\includegraphics[width=0.89\textwidth]{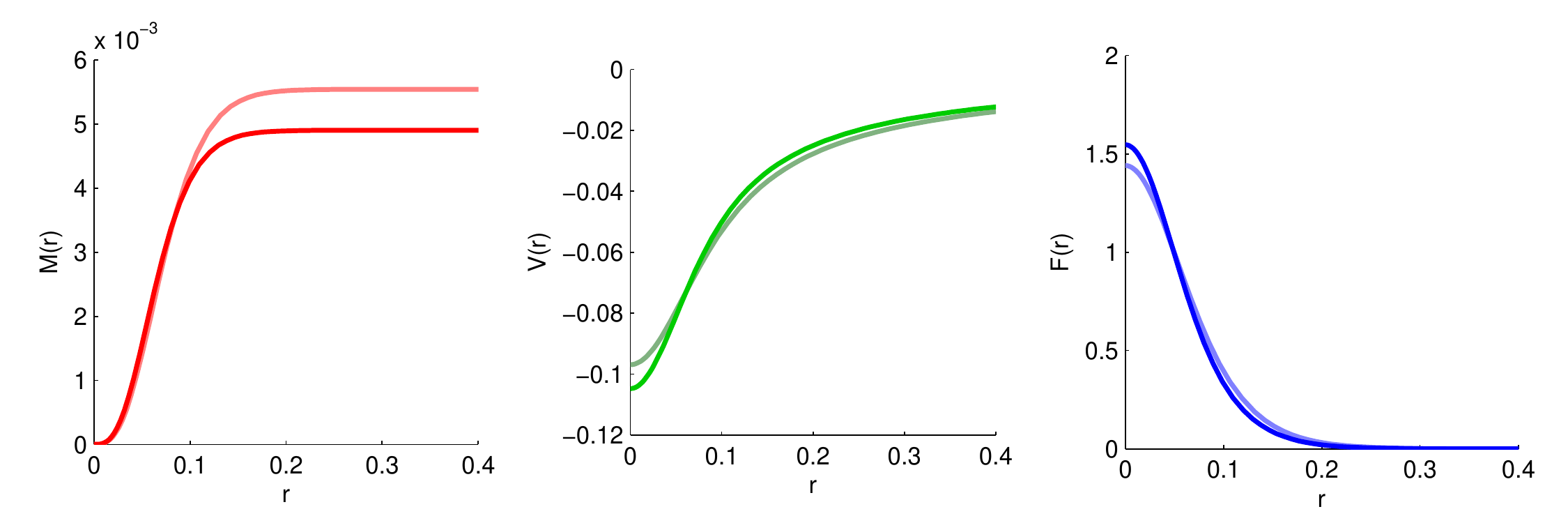}
\vspace{1pt}
\hrule
$\omega=90$\includegraphics[width=0.89\textwidth]{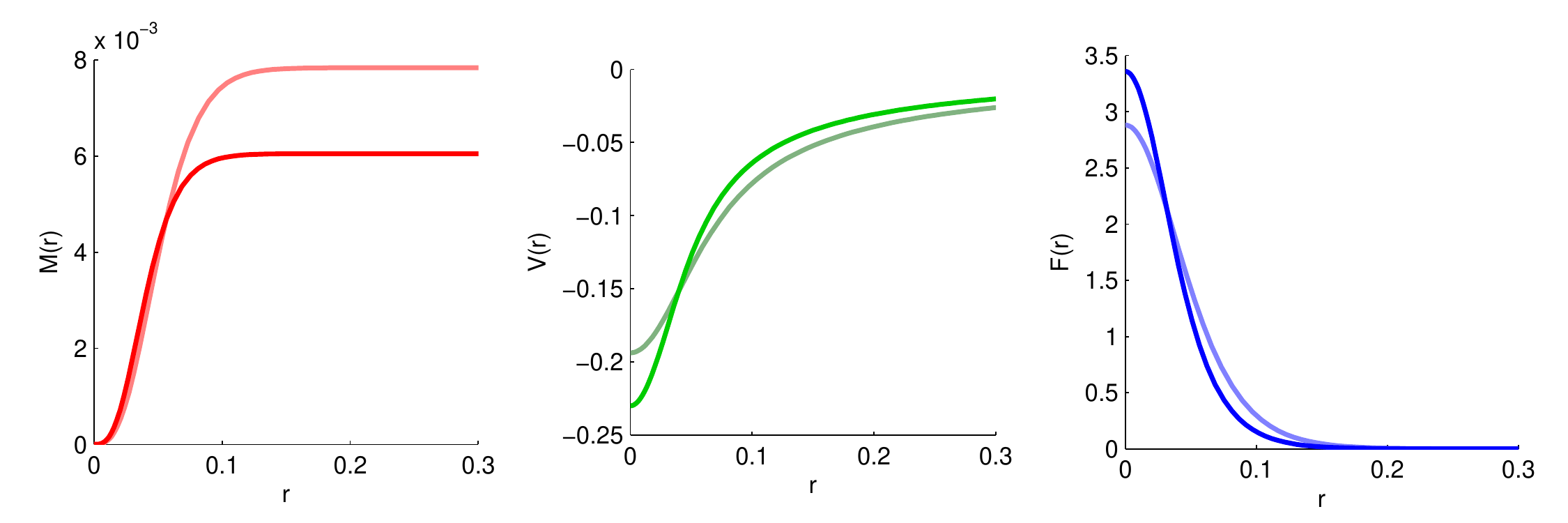}
\vspace{1pt}
\hrule
$\omega=85$\includegraphics[width=0.89\textwidth]{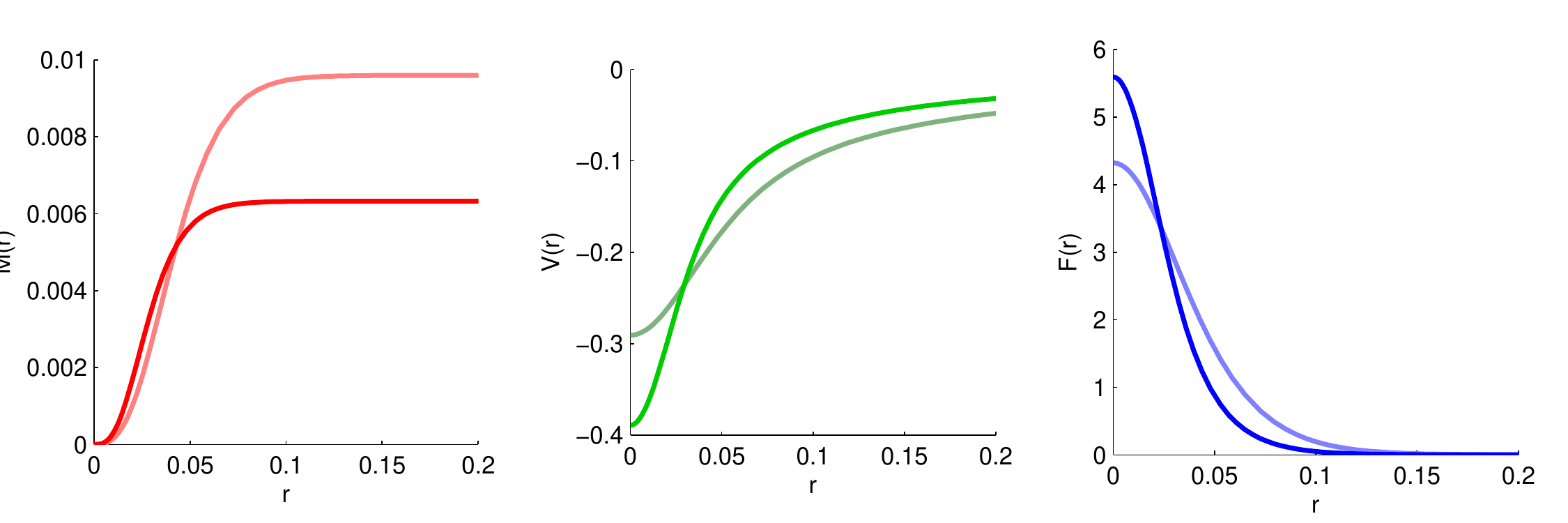}
\vspace{1pt}
\hrule
\caption{Solving for the ground state with $\Upsilon=100$ and $\omega=99,95,90,85$. As $\omega$ decreases we leave the low-field, nonrelativistic limit. The more solid graphs are the solutions obtained using the exact system \eqref{eq:ekgode}, and the fainter graphs are the solutions obtained using the low-field, nonrelativistic limit system \eqref{eq:ekgodelimit}.}
\label{fig:lowfieldlimit}
\end{figure}

\begin{definition} \label{def:rdm}
Given a static state $(\omega;M,V,F)$, we define $\RDM$ to be the radius at which the function $F$ switches from oscillatory to exponential behavior. We regard $r=\RDM$ as the edge of the dark matter halo because almost all the dark matter is contained in the region $r\leq\RDM$. See \cref{fig:RDM}.
\end{definition}
From \cref{eq:ekgodeF}, we see that $\RDM$ satisfies $\Upsilon^2-\omega^2 e^{-2V(\RDM)} = 0$, which implies
\begin{equation} \label{eq:rdmdef}
\omega e^{-V(\RDM)}=\Upsilon.
\end{equation}
This has a very natural physical interpretation---the frequency measured for the dark matter at $r=\RDM$ is always $\Upsilon$. See \cref{eq:otrdm}.

How do we manage to avoid the exponential growth term in the first solution in \cref{eq:ekgFsol}? A generic solution should include both terms, and indeed this is the case. However for special initial conditions, only the exponential decay term appears. This is why there are only countably many solutions for any fixed $\Upsilon$ and $\omega$. In fact our actual method of solving on the computer is as follows: fix a value for $V(0)$ consistent with $k(0)<0$ and then begin varying $F(0)$. Increasing $F(0)$ means that we get more oscillations before reaching $\RDM$ and decreasing $F(0)$ means fewer oscillations. Thus to get a ground state or a particular excited state we can only consider values for $F(0)$ in a particular range. In that range, there is only one value of $F(0)$ such that $F(r)$ exponentially decays for all $r>\RDM$. For all other values of $F(0)$, the solution includes an exponential growth term which dominates as $r\to\infty$, and we cannot consider such a solution because of the finite mass requirement. Once we have found (to within the desired accuracy) the correct value of $F(0)$, we then consider the fact that we probably do not have $V_\infty=0$. We then vary $V(0)$ and begin again. By varying $V(0)$ in an outer loop and $F(0)$ in an inner loop, we can find a solution $(\omega;M,V,F)$ satisfying all our requirements.

Next we consider the effect of changing $\omega$. Recall that $\Upsilon$ is to be regarded as a fundamental constant of nature. On the other hand, $\omega$ is a parameter we can vary. Once we have chosen $\omega$, there is then a unique ground state, first excited state, second excited state, etc. We find experimentally using the exact system \eqref{eq:ekgode} that for values of $\omega$ close to $\Upsilon$ we are in the low-field, nonrelativistic limit, but as $\omega$ decreases, the mass of the state increases, $\RDM$ decreases, and we leave the low-field, nonrelativistic limit. This is illustrated in \cref{fig:lowfieldlimit} for the ground state. Thus: \emph{For fixed $\Upsilon$, there is a one-parameter family of ground states (or first excited states, second excited states, etc.)} We can parametrize these states using the parameter $\omega$, but we can also parametrize them differently, for example, using $M_\infty$ or $\RDM$. In an upcoming paper \cite{goetz14} we will explore other ways to parametrize the states.

One interesting fact about this one-parameter family of states is that the states which fall in the low-field, nonrelativistic limit are all scalings of each other. Suppose $(\omega;M,V,F)$ solves the low-field, nonrelativistic limit system \eqref{eq:ekgodelimit}, and let $\lambda>0$ be a scaling factor. We can obtain another solution $(\bar{\omega};\bar{M},\bar{V},\bar{F})$ to \cref{eq:ekgodelimit} by performing the following scalings:
\begin{subequations} \label{eq:ekgscalings}
\begin{align}
\bar{r} &= \lambda^{-1} r \\
(1-\bar{\omega}/\Upsilon) &= \lambda^2(1-\omega/\Upsilon) \label{eq:ekgscalingsomega} \\
\bar{M} &= \lambda M \\
\bar{V} &= \lambda^2 V \\
\bar{F} &= \lambda^2 F.
\end{align}
\end{subequations}
This is easy to see directly from the system \eqref{eq:ekgodelimit}. Of course $\lambda$ cannot be too large or too small or we will leave the low-field, nonrelativistic limit. We can think of this state of affairs in the following way: in the low-field, nonrelativistic limit, there is a unique ground state (first excited state, second excited state, etc.) except for a scaling factor. Imposing one more condition will ``fix a scaling'' and give us a unique sequence of static states corresponding to $n=0,1,2,\ldots$.

\section{A Tully-Fisher-Like Relation} \label{sec:TF}
With this background we can describe the main result of this paper, which is that fixing the oscillation frequency of wave dark matter near the edge of dark galactic halos implies a Tully-Fisher-like relation for those halos. We will show this for the static states described above.

Given a finite mass static state $(\omega;M,V,F)$, we define
\begin{definitiondotless}
\begin{equation} \label{eq:omegatrue}
\omegatrue(r)=\omega e^{-V(r)}.
\end{equation}
\end{definitiondotless}
This is a physical quantity---the true frequency of the dark matter that would be measured at a particular value of $r$. The factor $e^{-V(r)}$ comes from the metric \eqref{eq:spacetime}. Using the previous definition and \cref{eq:rdmdef}, we see that $\RDM$ satisfies, and in fact is characterized by,
\begin{equation} \label{eq:otrdm}
\omegatrue(\RDM)=\Upsilon.
\end{equation}
Since $V(r)$ increases with $r$, $\omegatrue(r)$ decreases with $r$.

We now impose the condition
\begin{equation} \label{eq:sc}
\omegatrue(\RDM+r_0)=\omega_0
\end{equation}
for some fixed constants $r_0$ and $\omega_0$, where $r_0\ll\RDM$ and $\omega_0<\Upsilon$. It turns out that this is roughly equivalent to requiring the half-length of the exponential tail of each static state to be the same. This will be explored in detail in an upcoming paper \cite{goetz14}. Per our discussion in the previous section, this condition fixes a scaling of the static states. We wish to demonstrate that the sequence of static states ($n=0,1,2,\ldots$) obeys a Tully-Fisher-like relation.

\begin{figure}[bt]
\centering
\includegraphics[width=0.4\textwidth]{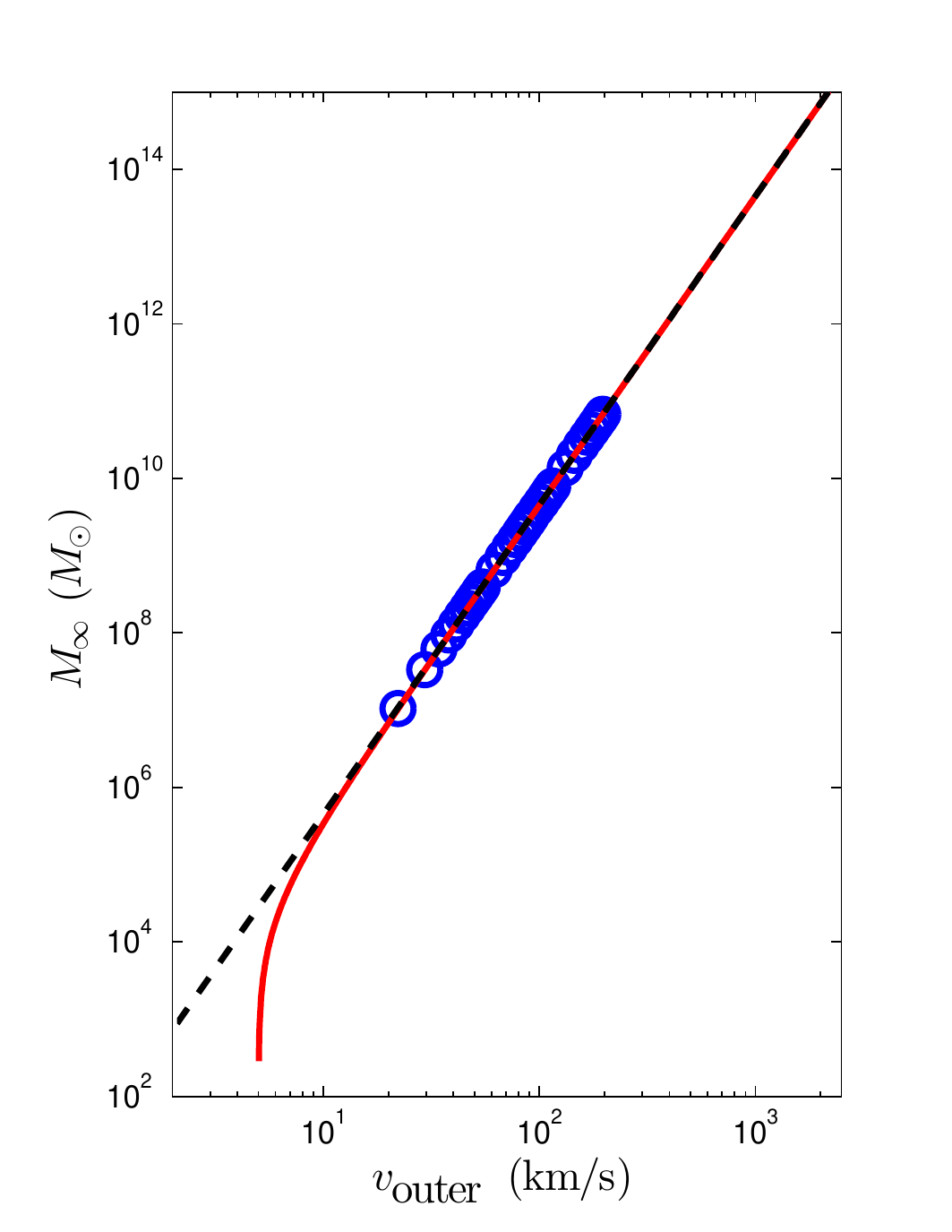} \includegraphics[width=0.4\textwidth]{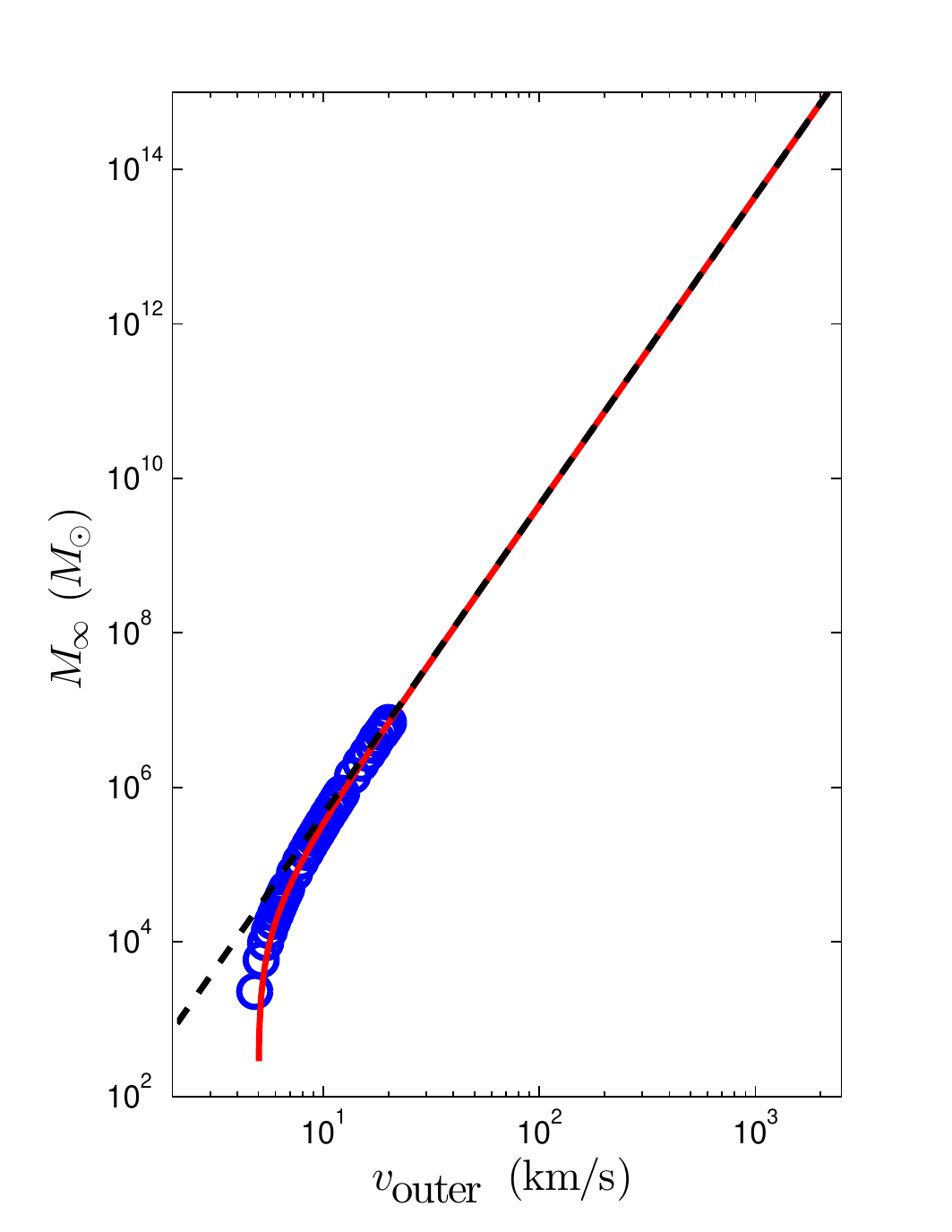}
\caption{A Tully-Fisher-like relation for static states obtained from solving the Einstein-Klein-Gordon equations \eqref{eq:ekg} using the system of ODEs \eqref{eq:ekgode}. These plots for $\Upsilon=100$ (left) and $\Upsilon=\num{100000}$ (right) show selected static states between $n=0$ and $n=500$ on the solid curve given by \cref{eq:loglog}. The ground state appears on the bottom left of each plot and the masses of the static states increase with $n$.}
\label{fig:tullyfishercurves}
\end{figure}

Let $(\omega;M,V,F)$ be one of these static states. For $r\geq\RDM$, $M(r)\approx M_\infty$ and thus from \cref{eq:ekgodeVlimit} we get $V(r)\approx -M_\infty/r$ for $r\geq\RDM$. Then from \cref{eq:omegatrue} we have
\begin{equation} \label{eq:rand}
\log\omegatrue(r)=\log\omega- V(r)\approx \log\omega+M_\infty/r
\end{equation}
for $r\geq\RDM$ so that, using \cref{eq:otrdm,eq:sc,eq:rand},
\begin{equation} \label{eq:tullyfisherderivation1}
\begin{split}
\frac{\log\Upsilon-\log\omega_0}{r_0} &= \frac{\log\omegatrue(\RDM)-\log\omegatrue(\RDM+r_0)}{r_0} \\
&\approx \frac{\frac{M_\infty}{\RDM}-\frac{M_\infty}{\RDM+r_0}}{r_0} \\
&= \frac{M_\infty}{\RDM(\RDM+r_0)}.
\end{split}
\end{equation}
Let $\vouter$ be the circular velocity of matter at $\RDM$. From Newtonian mechanics,
\begin{equation}
\vouter^2 = \text{acceleration}\cdot\RDM = \frac{M(\RDM)}{\RDM^2}\RDM \approx \frac{M_\infty}{\RDM}
\end{equation}
 Substituting $\RDM\approx M_\infty/\vouter^2$ into \cref{eq:tullyfisherderivation1}, we obtain
\begin{equation}
\frac{\log\Upsilon-\log\omega_0}{r_0} \approx \frac{M_\infty}{\frac{M_\infty}{\vouter^2}\left(\frac{M_\infty}{\vouter^2}+r_0\right)} = \frac{\vouter^4}{M_\infty+r_0\vouter^2}.
\end{equation}
This approximation holds for each static state $n=0,1,2,\ldots$. The expression on the left side is a constant. Looking at the right side, we see $\vouter^4$ in the numerator and $M_\infty$ in the denominator; thus, anticipating our final result below, we call the constant on the left side $\kTFtil^{-1}$. Thus
\begin{equation} \label{eq:tfconstdef}
\kTFtil^{-1} = \frac{\log\Upsilon-\log\omega_0}{r_0} \approx \frac{\vouter^4}{M_\infty+r_0\vouter^2}.
\end{equation}
Rearranging, we have
\begin{equation}
M_\infty\approx \kTFtil\vouter^4-r_0\vouter^2.
\end{equation}
To write this in a more palatable form, we introduce the constant
\begin{equation} \label{eq:vasydef}
\vasy^2=\frac{r_0}{\kTFtil}.
\end{equation}
We can then write
\begin{equation}
M_\infty \approx \kTFtil(\vouter^4-\vasy^2\vouter^2).
\end{equation}
Taking logarithms, we obtain
\begin{equation} \label{eq:loglog}
\log M_\infty \approx \log\kTFtil + 4\log\vouter + \log(1-(\vasy/\vouter)^2).
\end{equation}
This is our result. For $\vouter\gg\vasy$, we have
\begin{equation} \label{eq:loglog2}
\log M_\infty \approx \log\kTFtil + 4\log\vouter,
\end{equation}
i.e.
\begin{equation}
\frac{M_\infty}{\vouter^4} \approx\kTFtil.
\end{equation}
This is a Tully-Fisher-like relation. The log-log plot of $M_\infty$ versus $\vouter$ is a line with slope $4$. As $\vouter$ approaches $\vasy$ from above, the log-log plot exhibits a vertical asymptote at the velocity $\vasy$. See \cref{fig:tullyfishercurves}. It is interesting to note that the empirical baryonic Tully-Fisher relation also might exhibit this asymptote-like behavior---see, for example, Figure 1 of \cite{mcgaugh10}. On the other hand, other boundary conditions similar to \cref{eq:sc} do not give this vertical asymptote, as discussed in an upcoming paper \cite{goetz14}.

In \cref{fig:tullyfishercurves} we show two log-log plots modeled after Figure 1 of \cite{mcgaugh10} which show that the static states do indeed follow this Tully-Fisher-like relation. These static states were obtained by solving the exact system of ODEs \eqref{eq:ekgode}. To make these plots we needed to choose values for $\kTFtil$ and $\vasy$. For our purposes in this paper the specific values of these constants do not matter very much as we are only demonstrating the existence of a Tully-Fisher-like relation for wave dark matter and not yet trying to fit it to data. We chose $\kTFtil=\kTF$ where $\kTF$ is given by \cref{eq:tullyfisher} and $\vasy = \SI{5}{km/s}$. Using these values of $\vasy$ and $\kTFtil$, we used \cref{eq:tfconstdef,eq:vasydef} to solve for $r_0$ and $\omega_0$ and then had a computer generate the static states. We comment that the total mass $M_\infty$ is an increasing function of $n$.

The question of what $\kTFtil$ should be is an interesting one. The baryonic Tully-Fisher relation \eqref{eq:tullyfisher} is a relationship between total \emph{baryonic} matter $M_b$ and circular velocity $V$, where each galaxy's $V$ is obtained from a rotation curve which presumably does not extend all the way out to the edge of the dark matter halo. On the other hand, the Tully-Fisher-like relation which we have found for spherically symmetric static states is a relationship between the total \emph{dark} matter $M_\infty$ and the circular velocity $\vouter$ at $\RDM$. To connect these, we need to relate $M_b$ to $M_\infty$ and $V$ to $\vouter$. A reasonable guess is that they should be roughly proportional: $M_\infty \approx CM_b$ where $C>1$, since the quantity of dark matter dominates regular matter, and $\vouter\approx cV$, where $0<c<1$. In this case $\kTFtil$ would be related to $\kTF$ by 
\begin{equation}
\kTFtil = \frac{M_\infty}{\vouter^4} \approx \frac{C M_b}{(cV)^4} = \frac{C}{c^4}\kTF.
\end{equation}
Hence, to the extent that $\log(C)$ and $\log(c)$ are roughly constant across galaxies, then the baryonic Tully-Fisher relation is roughly equivalent to our Tully-Fisher-like relation for wave dark matter, as can be seen using \cref{eq:loglog2}.

\section{Discussion, Conjectures, and Testable Predictions} \label{sec:discussion}

One of the most important open questions in astrophysics concerns the nature of dark matter. In this section, we describe a testable prediction of the wave dark matter model (also known as scalar field dark matter (SFDM), boson star dark matter, and Bose-Einstein condensate (BEC) dark matter). Assuming a new idea called ``dark matter saturation'' described below and made precise in \cref{physconj:1}, given the distribution of the baryonic matter in a galaxy, we should be able to compute the distribution of the wave dark matter and hence the total mass distribution, as long as everything is approximately static and spherically symmetric. With the total mass distribution in hand it will be possible to compare to observations---for example, to compute rotation curves. These predictions will be possible once specific stability questions, described below, are answered.

If we assume that dark matter in many galaxies is approximately static and spherically symmetric, then it makes sense to look for static, spherically symmetric solutions to wave dark matter. Referring back to the scaling equations \eqref{eq:ekgscalings}, we see that there is a two-parameter family of static spherically symmetric solutions. These two parameters are $(n,\lambda)$, where $n \geq 0$ is an integer referring to the excited state and $\lambda$ is the scaling factor. Recall that $n=0$ refers to the ground state, and that more generally $n$ is the number of zeros of $F(r)$ from \cref{eq:staticstate}.

Now suppose that we solve for wave dark matter solutions in the presence of regular matter (which, for our purposes here, we will also assume is spherically symmetric and static). While the regular matter, through gravity, will change each wave dark matter solution, we still expect to find a two parameter $(n,\lambda)$ family of solutions.

One great benefit of our discussion so far is that it removes one of the parameters, namely the continuous parameter $\lambda$. In this paper, we showed that the boundary condition 
\begin{equation} \label{eq:boundarycondition}
\omegatrue(\RDM+r_0)=\omega_0,
\end{equation}
for some $r_0$ and $\omega_0$, is roughly what is needed to recover a Tully-Fisher-like relation. 
The boundary condition in \cref{eq:boundarycondition} is roughly equivalent to fixing the half-length of the exponentially decaying tail of the wave dark matter solution.  
In an upcoming paper \cite{goetz14}, \cref{eq:boundarycondition} is generalized to a class of ``Tully-Fisher boundary conditions'' which, generally speaking, fix some property of the wave dark matter near or at the edge of the halo. All Tully-Fisher boundary conditions are roughly equivalent and give a Tully-Fisher-like relation. To be clear, we have not explained why a Tully-Fisher boundary condition should be expected from the theory, just that something close to it seems necessary to be compatible with the observations which make up the baryonic Tully-Fisher relation. In any case, assuming this boundary condition effectively determines $\lambda$, leaving only one parameter free, namely $n$.

What should the value of $n$ be? Given a regular matter distribution, for each $n$ we get a precise wave dark matter distribution which satisfies the above boundary condition. Some of these wave dark matter solutions will be stable and some will be unstable. Numerical results show that ground states of wave dark matter are stable while excited states, without any other matter around, are unstable \cite{lai07}. On the other hand, \cite{lai07} also shows that excited states may be stabilized by the presence of another matter field.  Our conjecture is that the regular, visible, baryonic matter stabilizes wave dark matter in galaxies. In fact, given a regular matter distribution, we conjecture that there exists a largest value of $n$, call it $N$, for which the corresponding wave dark matter solution is stable.  We conjecture that galaxies are described best by choosing $n = N$.

The total mass of the spherically symmetric static states described in this paper increases with $n$ and we expect the same to be true for the distributions of dark matter we are describing now. Thus, setting $n = N$ is consistent with the idea that galaxies are ``dark matter saturated'', meaning that they are holding as much dark matter as possible, subject to the boundary condition above. Since galaxies typically exist in clusters which are mostly made of dark matter, it seems likely that they are regularly bombarded by dark matter, so that it would be natural for them to reach this state of saturation $n=N$.

To make this discussion precise, we need a model for the regular matter.  In order to study stability questions, we need to know how the regular matter distribution changes as the wave dark matter distribution changes, and vice versa. 

For example, a relatively simple way to model regular matter is with another scalar field.  There are others ways to model regular matter which we do not discuss here.  We caution the reader that this second scalar field is only a practical device for approximately modeling the regular baryonic matter---namely the gas, dust, and stars in a galaxy. In no way are we suggesting a second scalar field should exist physically. Furthermore, the parameters of this second scalar field are chosen simply to fit the regular matter distribution of a galaxy as well as possible.

Let $f_1$ exactly model wave dark matter with its fundamental constant of nature $\Upsilon_1$.  Let $f_2$ be a convenient device for approximately modeling the regular baryonic matter consisting of the gas, dust, and stars of a galaxy, where $\Upsilon_2$, which is not a fundamental constant of nature, is chosen as desired to best fit the regular matter.  The action is then  

\begin{equation}
\mathcal{F}(g,f_1,f_2) = \int \left[R_g-2\Lambda - 16\pi\left(\abs{f_1}^2 + \frac{\abs{df_1}^2}{\Upsilon_1^2} + \abs{f_2}^2 + \frac{\abs{df_2}^2}{\Upsilon_2^2} \right)\right]\,dV_g,
\end{equation}
where $\Lambda$ is the cosmological constant and may as well be assumed to be zero for our discussion on the scale of galaxies.  The above action results in the following Euler-Lagrange equations:
\begin{subequations} \label{eq:EL}
\begin{align}
\begin{split}
G+\Lambda g &= 8\pi\Bigg[ \frac{df_1\tensor d\bar{f}_1+d\bar{f}_1\tensor df_1}{\Upsilon_1^2}-\left(\abs{f_1}^2+\frac{\abs{df_1}^2}{\Upsilon_1^2}\right)g \\
& \qquad\quad \frac{df_2\tensor d\bar{f}_2+d\bar{f}_2\tensor df_2}{\Upsilon_2^2}-\left(\abs{f_2}^2+\frac{\abs{df_2}^2}{\Upsilon_2^2}\right)g \Bigg]
\end{split} \\
\Box_g f_1 &= \Upsilon_1^2 f_1 \\
\Box_g f_2 &= \Upsilon_2^2 f_2
\end{align}
\end{subequations}
We approximate the regular matter distribution with a ground state solution for $f_2$. We have two free parameters with which to approximate the given regular matter distribution, namely $\Upsilon_2$ and the ``scaling parameter'' for the ground state solution, which we could call $\lambda_2$. This should allow us to choose two physical characteristics of the regular matter. We choose to specify the total mass $M_b$ and the radius $R_b$ of the regular matter, perhaps defined as that radius within which some fixed percentage of the regular matter is contained.

As described already, we impose the boundary condition in \cref{eq:boundarycondition} for $f_1$. Then for each choice of $n \ge 0$, we get a solution to the system of equations \eqref{eq:EL} which reduces to a system of ODEs in a manner very similar as before. Some solutions will be stable and some will be unstable. Since we now have the dynamical equations \eqref{eq:EL}, these stability questions are now fairly well defined.  

Hence, for each $M_b$, $R_b$, and $n$, we get a static, spherically symmetric solution to \cref{eq:EL} satisfying the boundary condition in \cref{eq:boundarycondition}.  
\begin{mathconjecture}
In the low-field, nonrelativistic limit, for each choice of total regular mass $M_b$ and regular matter diameter $R_b$, there exists an integer $N \ge 0$ such that static, spherically symmetric solutions to \cref{eq:EL} satisfying the boundary condition in \cref{eq:boundarycondition} with $n \le N$ are stable and those with $n > N$ are unstable.   
\end{mathconjecture}
If this math conjecture is true, or even if there is just a largest or most massive stable $n$, then there is a natural physics conjecture to make as well.
\begin{physicsconjecture}[``Dark Matter Saturation''] \label{physconj:1}
The dark matter and total matter distributions of most galaxies which are approximately static and spherically symmetric are approximately described by static, spherically symmetric solutions to \cref{eq:EL} satisfying the boundary condition in \cref{eq:boundarycondition} with $n = N$.
\end{physicsconjecture}
This last conjecture only leaves three parameters open, namely $\Upsilon$, the fundamental constant of nature in the wave dark matter theory, and $r_0$ and $\omega_0$ from the boundary condition in \cref{eq:boundarycondition}. These last two parameters are equivalent to choosing values for $\kTFtil$ and $\vasy$, the last of which is not relevant for most galaxies. Hence, there are effectively only two parameters, namely $\Upsilon$ and $\kTFtil$, left open, with which to fit the dark matter and total matter distributions of most of the galaxies in the universe.  Hence, the physics conjecture stated above should be a good test of the wave dark matter theory.

\newpage

\printbibliography

\begin{table}[p]
\centering
\begin{tabular}{|l|c|c|c|c|}
\hline
\textbf{Unit} & \textbf{Seconds} & \textbf{Years} & \textbf{Meters} & \textbf{AUs} \\
\hline
\hline
kilogram & $\num{2.48e-36}$ & $\num{7.85e-44}$ & $\num{7.43e-28}$ & $\num{4.96e-39}$ \\
\hline
\hline
\textbf{Astronomical Body} & \textbf{Seconds} & \textbf{Years} & \textbf{Meters} & \textbf{AUs} \\
\hline
\hline
Sun & $\num{4.93e-6}$ & $\num{1.56e-13}$ & $\num{1480}$ & $\num{9.87e-9}$ \\
\hline
Earth & $\num{1.48e-11}$ & $\num{4.69e-19}$ & $\num{.00443}$ & $\num{2.96e-14}$ \\
\hline
Moon & $\num{1.82e-13}$ & $\num{5.77e-21}$ & $\num{5.45e-5}$ & $\num{3.65e-16}$ \\
\hline
Jupiter & $\num{4.70e-9}$ & $\num{1.49e-16}$ & $\num{1.41}$ & $\num{9.42e-12}$ \\
\hline
Cygnus X-1 black hole & $\num{7.4e-5}$ & $\num{2.3e-12}$ & $\num{22000}$ & $\num{1.5e-7}$ \\
\hline
Sag A* black hole & $\num{20}$ & $\num{6e-7}$ & $\num{6e9}$ & $\num{.04}$ \\
\hline
Milky Way & $\num{e7}$ & $\num{e-1}$ & $\num{e15}$ & $\num{e4}$ \\
\hline
\end{tabular}
\caption{Common masses in geometrical units of time or distance.} \label{table:masses}
\end{table}

\begin{table}[p]
\centering
\begin{tabular}{|l|c|c|c|c|}
\hline
\textbf{Unit} & \textbf{Meters} & \textbf{AUs} & \textbf{Kilograms} & \textbf{Solar Masses} \\
\hline
\hline
second & $\num{3.00e8}$ & $\num{.00200}$ & $\num{4.04e35}$ & $\num{203000}$ \\
\hline
day & $\num{2.59e13}$ & $\num{173}$ & $\num{3.49e40}$ & $\num{1.75e10}$ \\
\hline
year & $\num{9.46e15}$ & $\num{63200}$ & $\num{1.27e43}$ & $\num{6.41e12}$ \\
\hline
\hline
\textbf{Astronomical Time} & \textbf{Meters} & \textbf{AUs} & \textbf{Kilograms} & \textbf{Solar Masses} \\
\hline
\hline
age of universe & $\num{1.31e26}$ & $\num{8.73e14}$ & $\num{1.76e53}$ & $\num{8.84e22}$ \\
\hline
age of solar system & $\num{4.3e25}$ & $\num{2.9e14}$ & $\num{5.8e52}$ & $\num{2.9e22}$ \\
\hline
\end{tabular}
\caption{Common times in geometrical units of distance or mass.} \label{table:times}
\end{table}

\begin{table}[p]
\centering
\begin{tabular}{|l|c|c|c|c|}
\hline
\textbf{Unit} & \textbf{Kilograms} & \textbf{Solar Masses} & \textbf{Seconds} & \textbf{Years} \\
\hline
\hline
meter & $\num{1.35e27}$ & $\num{.000677}$ & $\num{3.34e-9}$ & $\num{1.06e-16}$ \\
\hline
AU & $\num{2.01e38}$ & $\num{1.01e8}$ & $\num{499}$ & $\num{1.58e-5}$ \\
\hline
light-year & $\num{1.27e43}$ & $\num{6.41e12}$ & $\num{3.16e7}$ & $\num{1}$ \\
\hline
parsec & $\num{4.16e43}$ & $\num{2.09e13}$ & $\num{1.03e8}$ & $\num{3.26}$ \\
\hline
\hline
\textbf{Astronomical Body} & \textbf{Kilograms} & \textbf{Solar Masses} & \textbf{Seconds} & \textbf{Years} \\
\hline
\hline
Sun (mean radius) & $\num{9.37e35}$ & $\num{471000}$ & $\num{2.32}$ & $\num{7.36e-8}$ \\
\hline
Earth (mean radius) & $\num{8.58e33}$ & $\num{4310}$ & $\num{.0213}$ & $\num{6.73e-10}$ \\
\hline
Moon (mean radius) & $\num{2.34e33}$ & $\num{1180}$ & $\num{.00579}$ & $\num{1.84e-10}$ \\
\hline
Jupiter (mean radius) & $\num{9.41e34}$ & $\num{47300}$ & $\num{.233}$ & $\num{7.39e-9}$ \\
\hline
\end{tabular}
\caption{Common distances in geometrical units of mass or time.} \label{table:distances}
\end{table}

\end{document}